\documentclass[sigconf,authorversion,nonacm]{acmart}
\AtBeginDocument{%
  \providecommand\BibTeX{{%
    \normalfont B\kern-0.5em{\scshape i\kern-0.25em b}\kern-0.8em\TeX}}}

\usepackage[most]{tcolorbox}

\usepackage{url}
\usepackage{color, colortbl}
\definecolor{Gray}{gray}{0.9}
\definecolor{LightCyan}{rgb}{0.88,1,1}
\usepackage{float}
\usepackage{makecell}

\usepackage{graphicx}
\usepackage{wrapfig}
\usepackage{nicematrix}

\usepackage{xcolor, soul}
\sethlcolor{LightCyan}

\usepackage{algorithm}
\usepackage{algpseudocode}

\newcommand{\QC}{{{\color{black}Partial-Query Checker}}}
\newcommand{\NQC}{{{\color{black}Neural Query Checker}}}
\newcommand{\SQC}{{{\color{black}Symbolic Query Checker}}}
\newcommand{\QTR}{{{\color{black}Query Test-and-Repair}}}

\renewcommand\footnotetextcopyrightpermission[1]{}

\title{Enhancing SQL Query Generation with Neurosymbolic Reasoning}

\begin{document}
\author{Henrijs Princis$^{1}$
\qquad
Cristina David$^{2}$ 
\qquad
Alan Mycroft$^{1}$}
\affiliation{$^{1}$University of Cambridge, Cambridge, UK 
\and
$^{2}$University of Bristol, Bristol, UK
}

\begin{abstract}
Neurosymbolic approaches blend the effectiveness of symbolic reasoning with the flexibility of neural networks. In this work, we propose a neurosymbolic architecture for generating SQL queries that builds and explores a solution tree using Best-First Search, with the possibility of backtracking. 
For this purpose, it integrates a Language Model (LM) with symbolic modules that help catch and correct errors made by the LM on SQL queries, as well as guiding the exploration of the solution tree.
We focus on improving the performance of smaller open-source LMs, and
we find that our tool, Xander,
increases accuracy by an average of 10.9\% and reduces runtime by an average of 28\%  compared to the LM without Xander, enabling a smaller LM (with Xander) to outperform its four-times-larger counterpart (without Xander).

\end{abstract}

\maketitle              

\section{Introduction}

Language models (LMs)\footnote{In this work, we use the term language model (LM) to refer to a language model of any size.} in particular \textit{large} language models (LLMs) have recently been successfully applied to a wide range of tasks including code generation~\cite{AlphaCode,Copilot,codeT5,CodeRL}. 
While initially LMs were used as black-box, monolithic, entities, recently there has been a shift towards architectures that display some form of logical reasoning,
where some defer parts of the reasoning to external modules~\cite{DBLP:journals/corr/abs-2205-00445}, or explore different paths when building the solution~\cite{DBLP:conf/nips/YaoYZS00N23}.

In this work, we follow this trend and propose an architecture for the generation of SQL queries that leverages external neurosymbolic reasoning to guide the (nonlinear) exploration of the solution space.
Our design is compatible with any pretrained LM, and we apply it to several smaller, open-source LMs, where we show that it improves their performance out-of-the-box. 
While our focus is on SQL, our broader goal is to illustrate that a neurosymbolic approach provides an alternative to scaling model size.
This addresses the high computational cost of LMs, where accuracy hinges heavily on the parameter count~\cite{ScalingLawsOfLargeLanguageModels}.

Our neurosymbolic 
architecture builds and explores a solution tree using Best-First Search~\cite{10.5555/525}, with the possibility of backtracking.
The exploration is guided 
by a symbolic query checker, a neural query checker and a symbolic test and repair module.
When integrated with a generic pretrained LM, these modules 
prune bad queries before they have finished generating and
correct errors made by the LM. 
As far as we are aware LM based nonlinear solution building has not been investigated in the context of code generation, where the solution space is particularly large.

One challenge when generating code is that the result can only be checked for correctness (i.e.\ adherence to the original intent) once it is completely generated. In order to address this, one of the innovations of our work is that the symbolic query checker 
is designed so that it is able to test incomplete queries. The ability to exclude erroneous candidates as early as possible is critical when exploring a vast solution space.

As an additional challenge, SQL's flexibility in expressing the same query in multiple ways can hinder the LM's performance, as the model is penalized for predicting logically equivalent but syntactically different queries.
To address this, we adapt NLP techniques~\cite{lowercasePreprocessing, preprocessingComparison} to eliminate unnecessary information from SQL queries before training an LM. In particular, we propose Normalized SQL, a standardized form that minimizes stylistic variations. Normalized SQL enhances both LM fine-tuning and inference, where it can be easily verified by symbolic modules. Our results show that Normalized SQL significantly improves overall LM accuracy.

\subsection{Query synthesis}
The task of query synthesis has been previously considered in literature either as \emph{text-to-SQL} or \emph{Query-by-Example (QBE)}: text-to-SQL denotes the task of generating SQL code from a textual description~\cite{parsingToTextToSQL, abstractLogicalForm, MITsolvesATIS,RESDSQL, RATSQL, RASAT, Picard}, whereas
QBE aims to find a query that returns user provided I/O examples \cite{MottinBook,QRE,s4QueryEnum,discoverQueriesEnumeration,QRE,SchyeEnum}.

In this work, we focus on the combined task of generating queries from natural language descriptions accompanied by I/O examples. Although accompanying natural language descriptions with I/O examples is commonly used for code generation in general-purpose languages~\cite{AlphaCode,CodeRL,DBLP:journals/corr/abs-2406-00515}, it  has been far less studied for SQL -- the only work we are aware of is~\cite{DuoQuest}. 

While approaches to text-to-SQL mostly rely on custom neural architectures, QBE techniques typically use symbolic reasoning. We explore a unified approach, where, instead of designing a custom neural model, we propose a neurosymbolic design that can be used to enhance any pretrained LM without modifications.

\subsection{Problem Definition}
\begin{figure}
\begin{center}
  \includegraphics[width=0.98\linewidth]{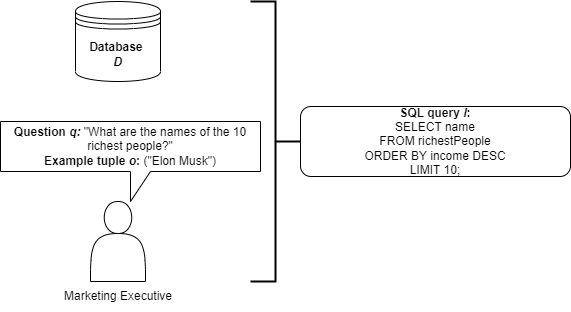}
  \caption{Illustration of the problem definition: given
  $(\mathcal{D},q,o)$ find $\ell$.
  }
  \label{fig:TaskDescription}
  \end{center}
\end{figure}

We consider the task of generating SQL code from a textual description and user-provided examples (see Figure~\ref{fig:TaskDescription}). We call this task text-to-SQL-with-examples and it can be defined as follows. Given a natural language question $q$, the database $\mathcal{D}$ which has a schema $\mathcal{D}_{schema}$, and possibly empty set of examples $o$, we wish to find an SQL query $l$ which when executed on database $\mathcal{D}$ answers question $q$ by returning a set $s$ of tuples such that $s \supseteq o$.  (This `open-world' formulation is
appropriate as we cannot expect the user to provide an
exhaustive set of examples $o$.)

Mathematically speaking, the text-to-SQL-with-examples task can be defined as finding a witness $\ell$ to the formula $(\exists \ell)\varphi(\ell,\mathcal{D},q,o)$ where $\varphi$ is a predicate that verifies that the query $\ell$ satisfies the natural language question $q$ and returns the right tuples $o$ on database $D$. We can write this more succinctly as \textbf{find}($\ell$ $\mid$ $\mathcal{D},q,o$)\footnote{Note natural language question $q$ could be ambiguous. We treat resolving English ambiguity as out-of-scope. We accept any query $\ell$ which satisfies the examples given.}.

\subsection{Contributions}
\begin{enumerate}
    \item We explore unifying symbolic and neural reasoning approaches for SQL query generation. In particular, we use LMs to generate SQL queries from natural language descriptions and examples, but guide the exploration of the solution space using two symbolic modules: (i) a 
    symbolic checker that verifies the correctness of incomplete queries, and
    (ii) a repair module which 
    uses \emph{fuzzing} to fix proposed-but-incorrect complete queries.
    
    \item We investigate the use of normalization in code generation by introducing \textit{Normalized SQL}\footnote{Normalized SQL refers to rewriting an SQL query into a specific form (see Section~\ref{sc:Normalized SQL}) and should not be confused with normal forms and relational database normalisation proposed by Codd.}, which increases LM's accuracy out of the box via finetuning, and can be easily verified during generation. 
    \item We implemented a prototype tool called Xander, and performed the following experiments:

\begin{enumerate}
\item Runtime and accuracy evaluation when using the neurosymbolic tool Xander with various open-source LMs.
    \item Ablation study on which parts of Xander bring the greatest improvements to accuracy and runtime. 
    \item Runtime and accuracy comparison between a neural and symbolic approach for detecting semantic, syntax and runtime query errors.
\end{enumerate}   
Xander was shown to lead to significant improvements
in runtime and accuracy, offering
a compelling alternative to scaling model size.
\end{enumerate}

\section{General Architecture of Xander}

\subsection{Our technique}
\begin{figure*}
\begin{center}
  \includegraphics[width=0.98\linewidth]{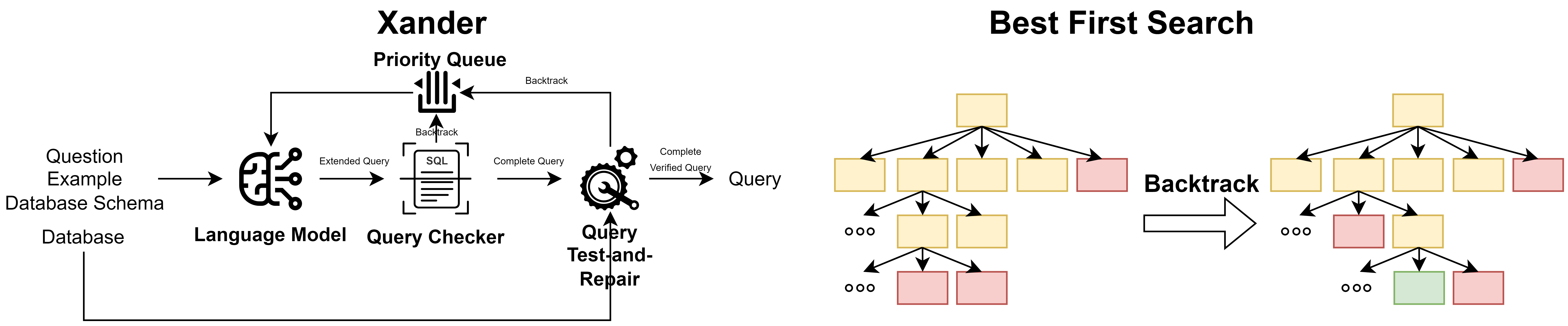}
  \caption{Xander overview. During inference Xander consists of three main modules connected by Best First Search (BFS). BFS explores the solution tree by choosing the most promising incomplete query (yellow box) and passing it to the LM.
  The LM then provides five candidates that extend the query by a single token. A \QC{} dismisses candidate queries containing errors (red box). The \QC{} passes queries without errors back to BFS if they are incomplete or to \QTR{} if they are complete. Finally, \QTR{} module executes the full query and verifies that it executes to the correct result, in which case it is returned to the user (green box). If it does not, they are dismissed (red box) and BFS backtracks.}
  \label{fig:TaskOverview}
  \end{center}
\end{figure*}

In this section, we present the design for our neurosymbolic technique, and its corresponding implementation Xander. A high level overview  of Xander during inference is given in Figure~\ref{fig:TaskOverview}, whereas a more detailed description is captured in Alg.~\ref{alg:sql}.

Xander takes as input a 3-tuple $(\mathcal{D},q,o)$ consisting of database $\mathcal{D}$, natural language question $q$, and example output tuples set $o$.
We then use $(\mathcal{D}_{schema},q,o)$, where $\mathcal{D}_{schema}$ is the schema of $D$ to prompt the language model.

In Alg.~\ref{alg:sql}, \textbf{x} denotes the task description consisting of the concatenation of database schema $\mathcal{D}_{schema}$, question $q$ and examples $o$, and $\textbf{y}$ represents the possibly empty partial query that will fulfill the task description once the query is fully generated. 

We use Byte-Pair Encoding (BPE)~\cite{BPE} to encode the task description and partial query as sequences of tokens meaning that one can think of \textbf{x} and \textbf{y} at lines 1 and 12 as: $\textbf{x} = [x^1,x^2,...,x^n]$ and $\textbf{y} = [y^1,y^2,...,y^{m-1}]$, respectively.

In each iteration of the while loop starting at line 4, we explore the current query $\ell$ (initially the empty string).
We first look at the scenario where query $\ell$ is not yet complete, meaning that we have to continue generating tokens for it. This is captured starting with line 10 in Alg.\ref{alg:sql}. After encoding $\ell$ into \textbf{y}, we call the LM to provide us the next token at line 12. 
Generally speaking, the LM defines a probability distribution $p$ over the possible continuations of $\textbf{y}$, where the most likely next token $y^m$ in the partial query is obtained with the following equation:
\begin{equation}
    y^m = \arg \max_{t} p(t| \mathbf{x}, \mathbf{y})
    \label{eq:LLMnextWord}
\end{equation}

In alg.~\ref{alg:sql}, the call to the LM returns the most promising five pairs of candidate next tokens and their respective probabilities, which get stored in $l\_continuations$. 

Each newly obtained query $\ell'$ (the concatenation of the previous partial query $\ell$ and the 
decoding of a newly generated token $t$)
are checked for plausibility by the \QC{} \textbf{PQC} (line 16 in alg.~\ref{alg:sql}).
As we explain later, the default \QC{} module applies simple symbolic checks to detect invalid queries.  Given that the LM generates queries one word at a time, completely generating a query and then checking it for errors is computationally wasteful. Instead, Xander checks queries that have only been partially generated, thus discarding many invalid queries early on.

The queries that pass the \QC{} are added to $priority\_queue$. This priority queue
allows us to avoid generating the same query twice, as well as to use Best First Search (BFS) in order to explore the space of candidate queries.
To enable the latter, the priority queue is ordered by the probability of each individual query. Then, BFS explores the solution tree by expanding the most promising node, which corresponds to the partial query with the highest probability.

For this purpose, in each iteration of the while loop at line 4, we pop the candidate partial query with the highest probability from the priority queue and further explore it (line 21 in alg.~\ref{alg:sql}).
Intuitively, the priority queue corresponds to a tree of possible candidates as shown on the RHS of Figure~\ref{fig:TaskOverview}, where each node represents a query (while some leaves may be complete queries, the rest are incomplete), and we can think of edges as corresponding to a token that concatenated to the parent query produces the child query. Notably, we can have cases where, for a given partial query, all the generated next tokens result in queries that have lower probability than some of the queries already in the priority queue. Consequently, the query that is popped in order to be explored in the next iteration of the while loop is one of the ancestors of the current query. This corresponds to backtracking in the tree of candidate queries.

If we find a complete query, it is checked by \QTR{} \textbf{QTR} (line 6 in alg.~\ref{alg:sql}) as described in alg.~\ref{alg:highlevelqueryrepairer}. Essentially, 
the query is executed on the full database $D$ to give a tuple set $s$.
If $s \supseteq o$, then the corresponding complete query is reported to the user as the proposed query.
Otherwise the \QTR{} module tries to repair $\ell$
using fuzzed (Hamming distance of one) variants $\ell'$ of $\ell$. Each $\ell'$ is executed and its tuple set compared with $o$ as before. The \texttt{HammingOneQueries} is a function that returns the set of queries that are a Hamming distance of 1 away from the original query. We regard any query that differs by exactly one aggregation operation (e.g. AVG), conditional operation (AND), comparison operator ($\geq$) or table column combination ($t1.c1$) as satisfying the requirement.

\begin{algorithm}
\caption{SQL Query Generation with Xander}\label{alg:sql}
\begin{algorithmic}[1]
\Require $\textbf{D}$: Database, $\textbf{q}$: question, $\textbf{o}$: example, $\textbf{BPE}$: Byte-pair encoding, $\textbf{BPD}$: Byte-pair decoding, $\textbf{LM}$: Language Model, 
$\textbf{PQC}$: \QC{}, $\textbf{QTR}$: \QTR{}
\State $\textbf{x} \gets \textbf{BPE}(D_{\text{schema}}, q, o)$ \Comment{The byte pair encoding of Database schema, question and example is used as input.}
\State \text{priority\_queue.setempty();}
\State $l, p_l \gets $ ``", 1
\While{\textbf{True}}
    \If{is\_complete\_query($l$)}
            \State $l \gets \textbf{QTR}(l, D)$  \Comment{Test and Attempt Repair}
            \If{\text{\textbf{QTR} succeeds}}
                \State \textbf{return} $l$
            \EndIf
    \Else
    \State $\textbf{y} \gets \textbf{BPE}(l)$
    \State $l\_continuations \gets \textbf{LM}(\textbf{x},\textbf{y})$ \Comment{Predict next tokens with probabilities}   
    \For{$t, p_t$ \textbf{in} $l\_continuations$}
        \State $l' \gets l + \textbf{BPD}(t)$\Comment{Proposed new query}
        \State $p_l' \gets p_l \times p_t$\Comment{Proposed probability}
        \If{\textbf{PQC}($l', o, D_{schema}$)}
            \State $\text{priority\_queue.add}((l',p_l'))$
        \EndIf
    \EndFor
    \EndIf
    \State $l, p_l \gets \text{priority\_queue.pop()}$ \Comment{Pop next $\ell$ to explore and its corresponding probability.}
\EndWhile
\end{algorithmic}
\end{algorithm}

\begin{algorithm}
\caption{High-Level \QTR{}}\label{alg:highlevelqueryrepairer}
\begin{algorithmic}[1]
\Require $l$: SQL query, $D$: Database
\Function{\textbf{QTR}}{$l$,$D$}
    \If{$o \subseteq$ execute(D, $l$)}
        \State \textbf{return success} $l$
    \EndIf
    \State 
    \Comment{Otherwise fuzz $\ell$ giving queries $\ell'$ to be tested}
    \For{$\ell'$ \textbf{in} \Call{HammingOneQueries}{$l$}}
        \If{$o \subseteq \text{execute}(D, \ell')$}
            \State \textbf{return success} $\ell'$
        \EndIf
    \EndFor
    \State \textbf{return fail}
\EndFunction
\end{algorithmic}
\end{algorithm}

The output of Xander is Normalized SQL which is a modified form of SQL that enforces stricter syntax and a consistent style. Normalized SQL increases accuracy and reduces the difficulty of checking whether an incomplete query contains an error (more details in the following section).

In Section~\ref{sc:Normalized SQL} we introduce Normalized SQL, in Section~\ref{sc:QC} we describe the \QC{} component
(and its symbolic and neural alternatives)
and finally in Section~\ref{sc:QTR} we justify the \QTR{} module.

\subsection{Normalized SQL}
\label{sc:Normalized SQL}

In this section we outline the limitations of SQL in the context of LMs, and overcome them using Normalized SQL.

In SQL, there are many different ways of expressing the same query.
A training dataset naturally may use varying styles, but
this hurts the performance of LMs trained on such SQL datasets.
This is because during training the LM is penalized for predicting a logically, but not lexically, equivalent query to the gold-standard query. Since different humans use different styles of writing queries, the LM will be randomly penalised for getting the query right, but using a different style (e.g.\ by capitalising instead of lower-casing SQL keywords). Therefore, training a LM on a dataset that contains mixed styles of SQL produces a similar effect as adding random noise during training.

This problem is not exclusive to SQL. For instance, it has been explored in detail in natural language processing, where several works 
suggest eliminating unnecessary information before training a LM.
This is achieved with data pre-processing methods such as punctuation removal, stop word removal, stemming, and lower casing ~\cite{lowercasePreprocessing, preprocessingComparison}. In SQL, existing works explore alternative model architectures which make the SQL produced correct by construction ~\cite{RLonSQL, SyntaxSQLNet}. 

\begin{figure}[ht]
\begin{center}
  \includegraphics[width=0.98\linewidth]{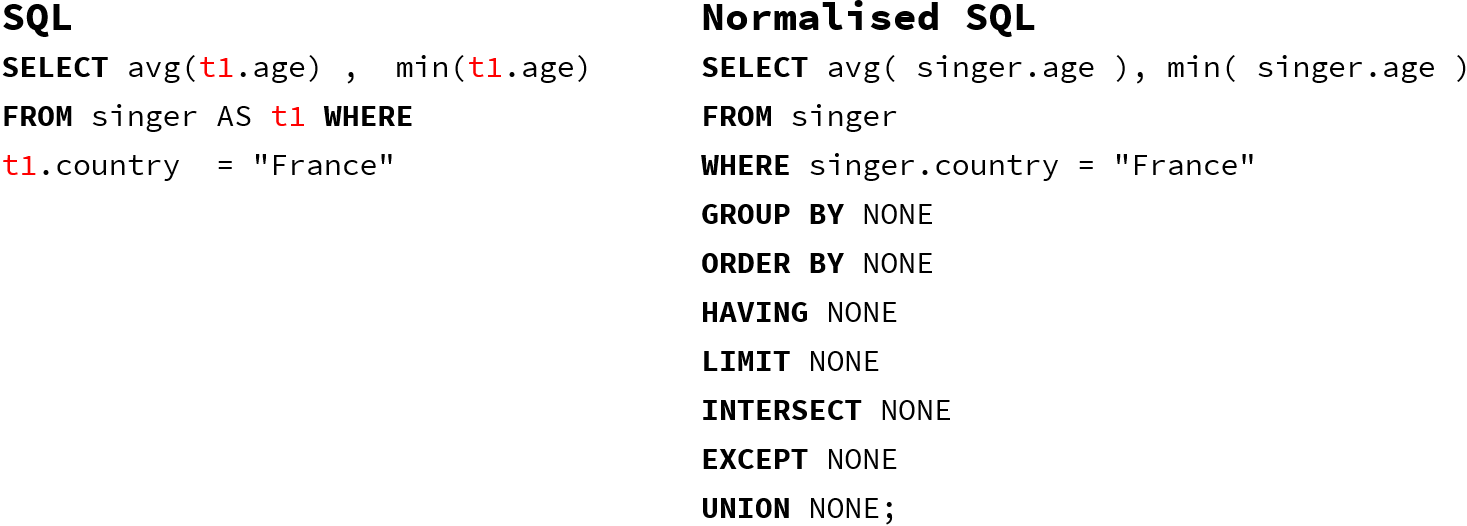}
  \caption{SQL (left) Normalized SQL (right). SQL uses an alias which is highlighted in red. Normalized SQL replaces the alias with what it points to (in this example, t1 is replaced with singer).}
  \label{fig:SimplifiedSQL}
  \end{center}
\end{figure}

\begin{figure}[ht!]
\begin{center}
  \includegraphics[width=0.98\linewidth]{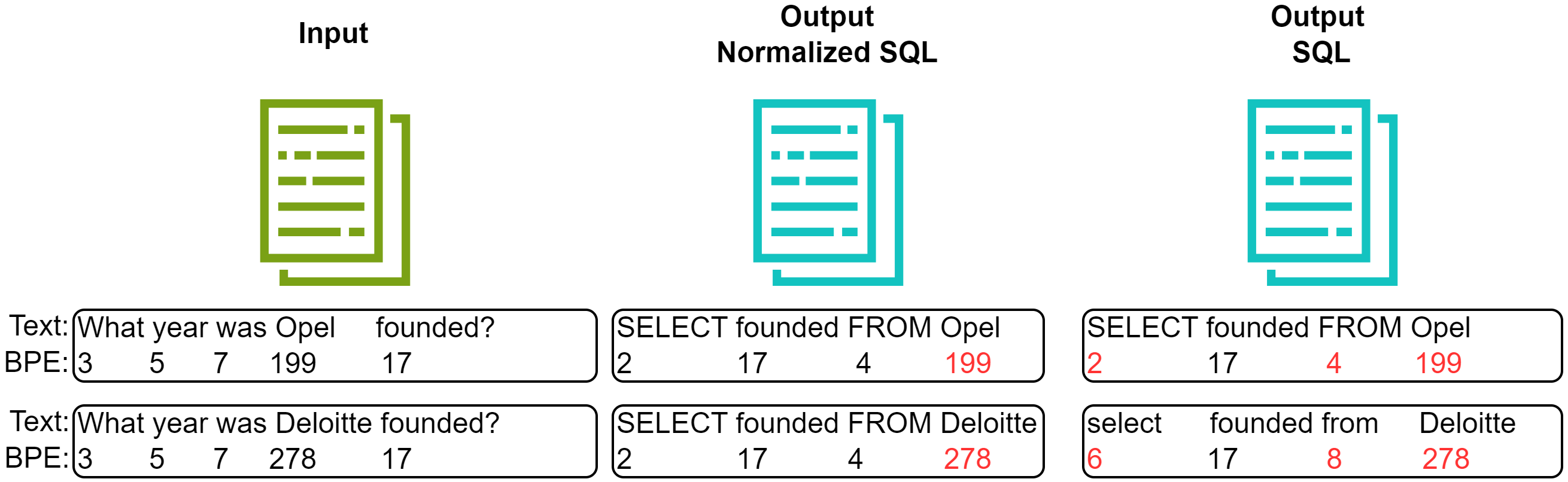}
  \caption{Illustration of how Normalized SQL makes learning easier. Notice how in (unnormalized) SQL case, despite a minor change to the input specification, a drastic change is observed in the output's byte-pair encoding.}
  \label{fig:NormSQL}
  \end{center}
\end{figure}

Following this direction, we designed  Normalized SQL~-- 
a modified form of SQL enforcing a stricter syntax that limits the number of ways of expressing the same query. 
In particular Normalized SQL uses the following style:
\begin{itemize}
    \item {We follow SQL typing convention and uppercase all SQL keywords.}
    \item {Each keyword is followed by a space. This choice makes vocabulary checking easier.}
    \item {Each clause is a single line long. This choice makes checking clauses for correctness easier.}
    \item {All clauses are included and non-optional. This is to eliminate the need for the LM to learn the clause order (e.g. a query where FROM precedes SELECT will error).}
    \item {We expand all aliases because during testing, we discovered that language models frequently struggle to determine the association between columns and tables when aliases are used in queries.\footnote{Unfortunately, this change means that complex join queries are not possible in Normalized SQL. These are uncommon, for example, in the Spider\cite{Spider} dataset only 1.5\% of queries use complex joins.}.}
\end{itemize}
An example of an SQL query and its corresponding Normalized SQL is given in Figure~\ref{fig:SimplifiedSQL}. Then, in Figure~\ref{fig:NormSQL}, we have two input natural language questions: ``What year was Opel founded?'' and ``What year was Deloitte founded?'', and their corresponding SQL and Normalized SQL output queries. The figure also contains the byte-pair encodings. For Normalized SQL, the difference in the output's encoding between the two queries is considerably smaller than for SQL. The variation in SQL comes down to low-level stylistic choices chosen by different annotators which makes the BPE output look artificially more different than it is.

One of the benefits of Normalized SQL is that it enables the checking of partial queries by the \QC{}. Intuitively, the stricter syntax of Normalized SQL makes it easier to find faulty queries early on (more details in the next section).

\subsection{\QC{}}
\label{sc:QC}

\begin{figure}[ht]
\begin{center}
  \includegraphics[width=0.98\linewidth]{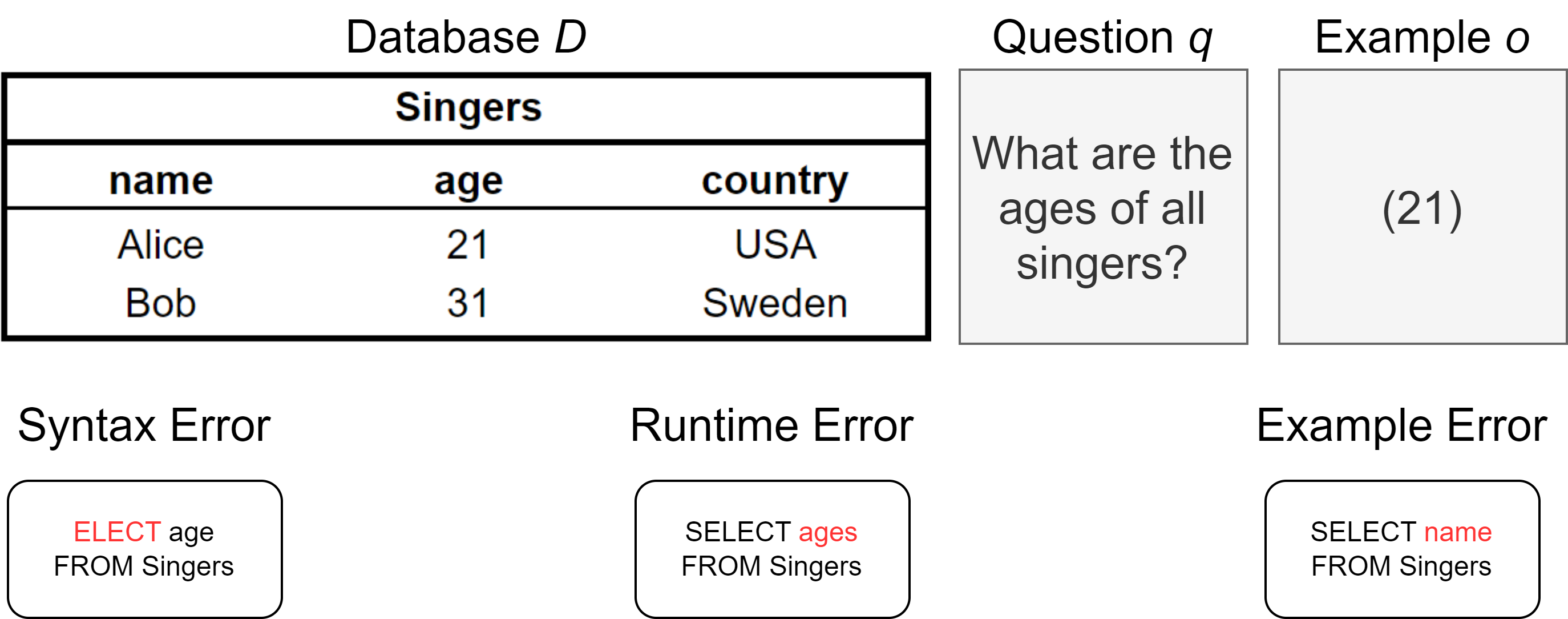} 
\end{center}
  \caption{Illustration of three different types of errors. Notice how example error query is syntactically correct and executes, but the tuples brought back do not contain the tuple provided by the user.}
\label{fig:example_errors}
\end{figure}

An important part of Xander's architecture is the \QC{}, which takes as input the initial specification (i.e. the question in natural language and the output example), the database schema and a potentially incomplete
SQL query $\mathbf{y}$, and investigates whether the query is valid (i.e. it can be completed such that the final query corresponds to the initial question) or invalid (i.e. there is no way of completing it in a way that corresponds to the initial question). For this purpose, the
\QC{} checks the query for common syntax, runtime, and example errors (see Figure~\ref{fig:example_errors} for instances of such errors).
An \emph{example error} occurs when the generated query does not return the output example provided by the user as part of the initial specification.

At first it is surprising that example errors can be detected without
running the query on the database, but sometimes (often enough)
example errors can be detected just given the database schema --
using techniques described below.
As aforementioned, the \QC{} only has access to the database schema
and not the full database (by contrast the \QTR{} module described in Section~\ref{sc:QTR} tests queries
on the full database).

As part of our evaluation, the \QC{} is pluggable -- we have
a standard \QC{}, the \emph{\SQC{}}, but also
a neural alternative, the \emph{\NQC{}}.
Both are explained below.

\subsubsection{\SQC{}}
\label{sc:SQC}
In order to detect the aforementioned errors, the \QC{} performs three types of checks: vocabulary, scope and type checking.

\paragraph*{Vocabulary Checking} The Normalized SQL query is checked clause by clause. Every clause is checked for containing the right vocabulary. For example, the \textbf{SELECT} clause can only contain aggregation operations, brackets and table-column names. In this step we also reject table-column combinations which do not belong to the schema $\mathcal{D}_{schema}$. Most errors caught by vocabulary checking will be syntax errors; however, it will also catch the runtime error of selecting an invalid table-column combination.

\paragraph*{Scope Checking} We check that all of the tables referred to by \textbf{SELECT}, \textbf{WHERE}, \textbf{GROUPBY}, and \textbf{HAVING}, clauses also appear in the \textbf{FROM} clause. Scope checking can only detect runtime errors. 

\paragraph*{Type Checking} Finally, the types of the tuples which are returned by the \textbf{SELECT} clause are compared to the type of the tuple provided as an example. This ensures that the right number of columns with the right types are chosen. Type checking only catches example errors. For illustration, in the example in Figure~\ref{fig:example_errors}, the query used to illustrate the example error returns a string, whereas the expected output is an integer.

\subsubsection{\NQC{}}
\label{sc:NQC}

We now explore an alternative plug-in to the \SQC{} described above. As part of the experimental evaluation, we conducted an experiment comparing the two query checkers, which concluded that the \SQC{} is more accurate than the \NQC{}, and thus, that's the one we use in our experiments. However, for completeness reasons, we also discuss the \NQC{}.

The \NQC{} is a neural network which aims to classify whether a possibly incomplete query is correct or contains an error. Such an error-detecting network is referred to as a \textit{critic} in the context of reinforcement learning.

\begin{figure}
\begin{center}
  \includegraphics[width=0.98\linewidth]{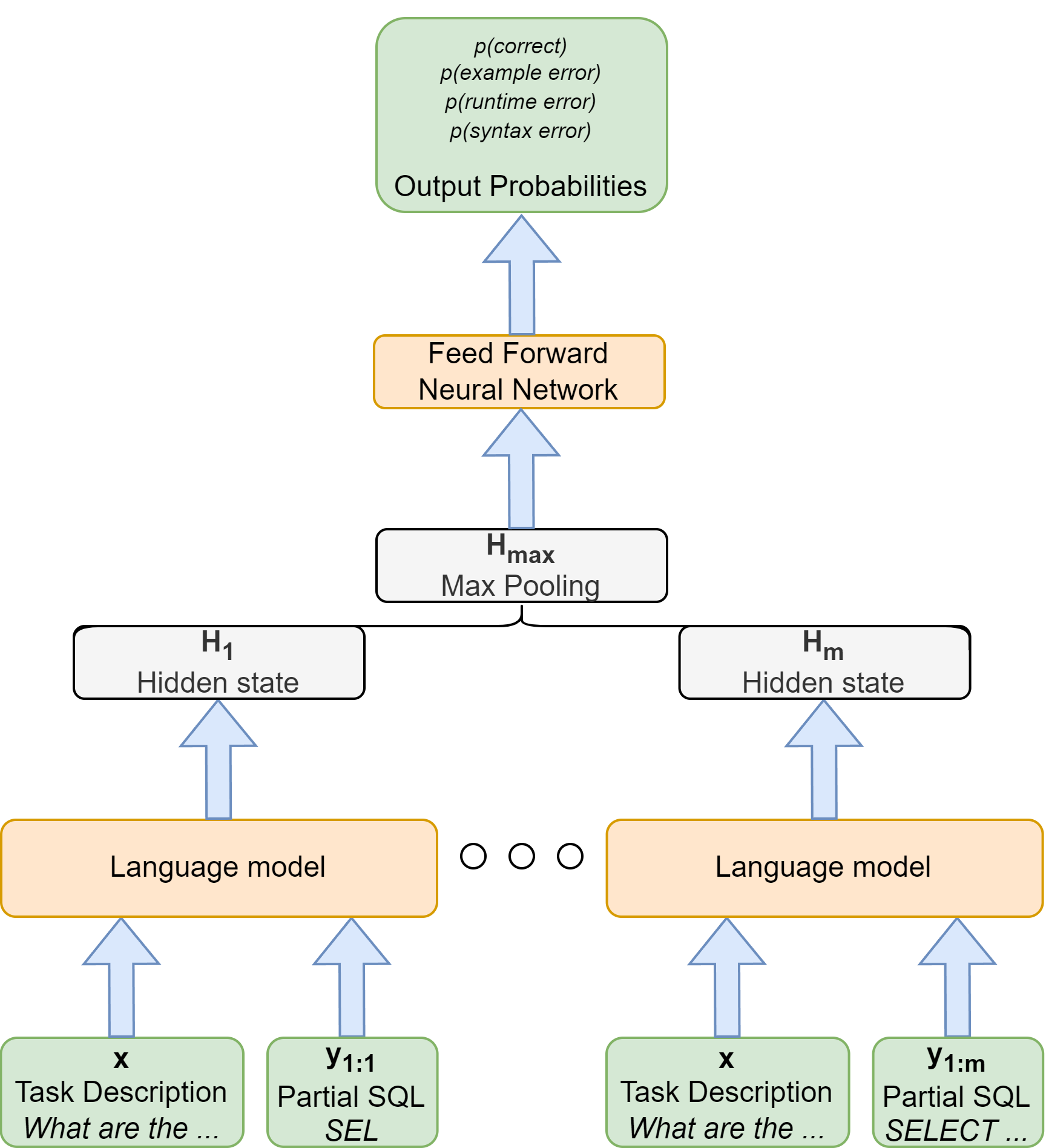}  
\end{center}
  \caption{Illustration of the \NQC{}. $\mathbf{y_{1:i}}$ denotes the subsequence of the tokenized partial Normalized SQL query generated by a language model up to token $i$. The length of the partial query is $m$. A hidden state $\mathbf{H_{i}}$ is produced for every subsequence starting at position 1. This hidden state $\mathbf{H_{i}}$ is then used as input to max pooling which works across sequence length. The result is input into another Feed Forward Neural Network and finally Softmax is applied to generate the output probabilities which correspond to `Correct', `Example error', `Runtime error', and `Syntax error'.}
\label{fig:TransformerQueryChecker}
\end{figure}

The main challenge in classifying queries using a neural network is that we do not know where the error happened. 
To overcome this challenge, we adopt a solution introduced in CodeRL~\cite{CodeRL} (see Figure~\ref{fig:TransformerQueryChecker}). Intuitively, the generated query is unveiled a word at a time to the \NQC{}. For every new word the \NQC{} sees, it outputs a hidden state containing information about whether a mistake has been made so far. Once we have gathered a hidden state for each word, we perform max pooling to obtain information about whether the entire Normalized SQL sequence contains a mistake at any point.

To train the \NQC{}, we generate 
queries using the training dataset, one of the finetuned LMs and Beam Search~\cite{Beamsearch}, which uses less memory then Best First Search. The aim is to mimic the distribution of queries we will see during inference. We execute the queries and assign them one of the following labels `Correct' when the query executes to the gold standard result, `Example error' when the query executes to a different result, `Runtime error' when the query is syntactically correct, but does not execute on a given database, and `Syntax error' when the query fails to run on any database. 

Notably, during training, we only use complete queries. 
We assume that due to the network's structure, in particular, the maximum pooling layer across sequence length dimension (see Figure \ref{fig:TransformerQueryChecker}), its accuracy on complete queries will be similar to partial queries, i.e.\ it will be able to identify mistakes as soon as they happen. 

\subsection{\QTR{}}
\label{sc:QTR}

\begin{figure}[ht]
\begin{center}
  \includegraphics[width=0.98\linewidth]{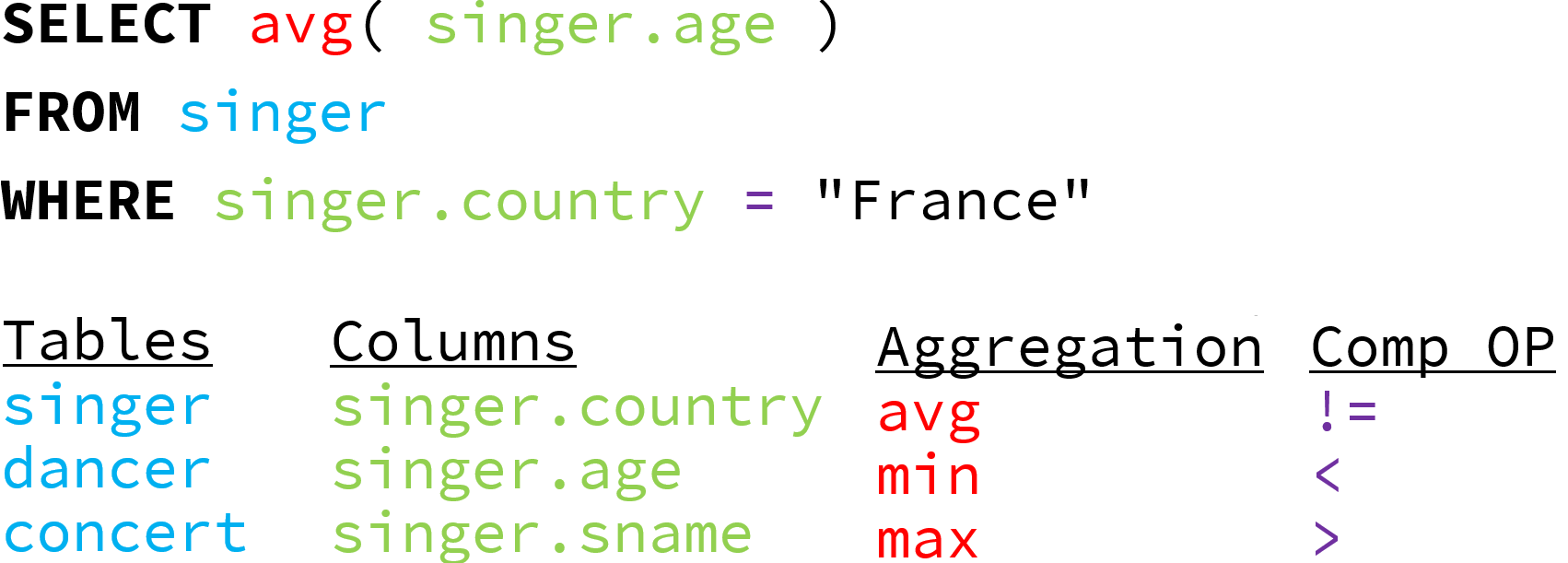} 
\end{center}
  \caption{Illustration of \QTR{} Module. It takes in a given query (top of figure) and then generates all queries that differ by one aggregation (red), table name (blue), column name (green), or comparison operator (purple).  An example query it would generate is \textbf{SELECT} \textit{max}(singer.age) \textbf{FROM} singer \textbf{WHERE} singer.country = ``France", which differs from the given query in that it uses ``max'' instead of ``avg'' (the Hamming distance between the two queries is 1).
   }
\label{fig:EnumerativeQueryChecker}
\end{figure}

Given a complete query, after we checked it with the \QC{}, we execute it to verify that it satisfies the user-provided example. If it does not, we attempt to repair it.
The \QTR{} module achieves this (as described in Alg.~\ref{alg:highlevelqueryrepairer}).

An important design consideration of the \QTR{} module is that it should keep the (Hamming) difference between the repaired query and the generated query to a minimum. If the repaired query diverges significantly from the original, it may indicate a flawed correction that overlooks the user's language specification, resulting in a potential false positive.

In this work the \QTR{} module generates all queries that differ from the original by exactly one \emph{enumerative} SQL token
(these are SQL operations, column names, or table names).
An example is given in Figure~\ref{fig:EnumerativeQueryChecker}.
It does not try to repair off-by-one errors for constants (e.g.\ by changing ``France" to ``Grance") due to very large search-space and low likelihood of success.

\section{Experiments}

We perform three experiments. First, we explore how much more efficient a neurosymbolic approach is compared to a purely neural approach. To answer this question we compare the runtime and performance of using a wide range of popular LMs on their own and then as a part of the neurosymbolic tool Xander. We call this experiment \textbf{generalisability} since it measures the generalisability of Xander over various LMs. We report the results of the generalisability experiment in Section~ \ref{sc:Generalisability}.

Next, we investigate what are the most important factors for determining Xander's performance. To answer this question we run an \textbf{ablation} experiment. We measure the accuracy and runtime benefits of Normalized SQL, \QC{}, \QTR{}, and whether or not examples are provided as a part of the input specification.
We report the results of this experiment in Section~\ref{sc:Xander Ablation}.

The \textbf{\QC{}} Comparison Experiment aims to answer whether we could improve Xander by replacing the \SQC{}  with \NQC{}. The results of this experiment are reported in Section~\ref{sc:QCCompareExperiment}.


\subsection{Experimental Setup}
\label{sc:Experimental Setup}
\paragraph*{Environment} For all experiments we used Python with the Hugging Face transformers library~\cite{HuggingFace}. All experiments except those for Microsoft Phi-1.5 were performed using Tesla P100 GPU and Intel Xeon 6142 CPU. For Microsoft Phi-1.5, due to the larger model size, we used an Amazon EC2 G5.xlarge instance with an A10G (24GB) GPU. 

\paragraph*{LLMs}
We considered the following LLMs for which weights are publicly available:
CodeT5~\cite{codeT5}, BART~\cite{BART}, CodeGen~\cite{codegen}, and Microsoft Phi-1.5 \cite{MicrosoftPhi}. As a  comparison point, we also used ChatGPTv3.5.

\paragraph*{Dataset}
We used the Spider~\cite{Spider} dataset in all experiments. The training set was used to finetune the network and the validation set was used to measure real-world accuracy and runtime. Since Spider dataset does not include examples in the user description, we generate them by executing the golden queries and taking the first returned tuple.

As we work with Normalized SQL, we rewrote the Spider dataset to follow Normalized SQL's constraints. There are a few queries in the original Spider dataset that use self joins and, therefore, cannot be expressed in Normalized SQL. The total number of queries in the original and the Normalized SQL dataset are provided in 
Table~\ref{table:spider}.

\begin{table}[h]
\centering
\begin{tabular}{ |c|c|c| } 
 \hline
 & \makecell{Training Dataset \\ \# Queries} & \makecell{Validation Dataset \\ \# Queries} \\ 
 \hline
 original & 7000 & 1034 \\ 
 \hline
 Normalized SQL & 6779 & 1018 \\ 
 \hline
\end{tabular}
\caption{Original Spider dataset vs. the one rewritten to follow the Normalized SQL constraints. }
\label{table:spider}
\end{table}

\paragraph*{Training} 
Throughout our experiments we use pretrained LMs.
 These LMs are further trained (finetuned) to generate SQL or Normalized SQL on the Spider dataset or its Normalized SQL version, depending on the experiment. 
All networks except Microsoft Phi-1.5 were fitted for 50 epochs with batch size of 10. The Adam~\cite{Adam} optimizer with a learning rate of $4e^{-5}$ was used to find the optimal weights. 

For Microsoft Phi-1.5, 
to save memory, batchsize of 1 was used and RMSProp optimizer was used instead of Adam. To account for the larger network size, the learning rate was lowered to 4e$^{-6}$ and the network was fitted for 5 epochs.

\paragraph*{Encoding} The input to the transformer was the byte-pair encoding $\mathit{BPE}$ of the serialised concatenation of the user question, database schema, and examples, i.e.\ \newline $\mathit{BPE}(Concat(str(q),str(\mathcal{D}_{schema}),str(o)),32100)$
where $32100$ is the size of the vocabulary. The database contents were not included in the specification because doing so would require too much memory.
The golden output was the tokenization of SQL or Normalized SQL depending on the experiment, both for training and validation datasets. 

\paragraph*{Query Evaluation} Queries were evaluated using Distilled Test Suites~\cite{evalProcedure} which provides exact match accuracy and execution accuracy. Exact match accuracy checks whether the abstract syntax tree of both queries is the same (excluding constants) whereas execution accuracy only checks whether each query returns the right tuples on a specific database instance. 

\subsection{Generalisability}
\label{sc:Generalisability}
The aim of this experiment is to explore whether a neurosymbolic approach is faster in terms of runtime and accuracy compared to a purely neural approach. The secondary goal of the experiment is to investigate whether a neurosymbolic approach could provide similar accuracy benefits as scaling the model size. 

In all instances, a one-minute time limit was given to generate a query which returns the user-provided output tuple when executed. In order to enable a fair comparison with Xander, the approach without Xander also constructs the solution by querying the LM for one token at a time and building a solution tree that gets explored using BFS. Essentially, it follows the same algorithm as the one for Xander in Alg.~\ref{alg:sql}, with the only exceptions that, at line 6, it only tests the complete query without attempting repair, and it does not call the \QC{} at line 16. Instead, all partial queries are added to the priority queue. For ChatGPT, 
we accessed it through OpenAI's API 
using a single query due to monetary limitations.

The results of this experiment can be seen in Table \ref{table:generalisability}, where Validation Exact
Match Accuracy refers to the percentage of generated queries that are a perfect match to the golden query, whereas Validation Execution Accuracy refers to the percentage of queries that, when run, produce the correct result.

\begin{tcolorbox}[colback=gray!7, colframe=gray!7, boxrule=0pt, left=5pt, right=5pt, top=5pt, bottom=5pt,breakable]
\begin{itemize}
\item[$\bullet$] {First, we see that \textbf{the neurosymbolic tool Xander increases the validation execution accuracy by an average of 10.9\%{\footnote{10.9\% is obtained by subtracting the average accuracy of all networks with Xander from the average execution accuracy without Xander}},
validation exact match accuracy by 6.0\% and it finds the correct query 28\%{\footnote{28\% is obtained by averaging each network's relative performance increase with Xander.}}} faster than using a purely neural approach in the form of a LM. 

Xander always improves execution accuracy, but we note that for BART model it decreased exact match accuracy. This is likely because exact match accuracy has a high false negative rate.} 

\item[$\bullet$] {Second, we see that a neurosymbolic approach offers a compelling alternative to scaling model size. \textbf{The CodeT5 small model \textit{beats} the four-times-larger CodeT5 base model when CodeT5 small is used with Xander.} Also on this point, 
ChatGPTv3.5 performed worse than up to an order of magnitude smaller fine-tuned LMs with Xander.}

\end{itemize}
\end{tcolorbox}

To test our hypothesis that our neurosymbolic methods improve the performance of smaller, open-source, models, we compare it against ChatGPT3.5.

\begin{table*}
\begin{center}
\renewcommand{\arraystretch}{1.2}
\begin{NiceTabular}{l r c c c c}
\textbf{Model} & \#parameters & \makecell{Training Time} &  \makecell{Validation Time \\ per sample (s)} & \makecell {Validation Exact\\ Match Accuracy (\%)} & \makecell{Validation Execution \\ Accuracy (\%)} \\
\hline
\textbf{CodeT5 small} & (62M) & 3h8m & 15.7 & 60.7 & 69.9 \\
\qquad \textbf{Ditto with Xander}&& 2h59m & 9.1 & 66.0 & 78.6\\
\hline
\textbf{{CodeT5 base}} & (222M) & 10h8m & 10.7 & 63.5 & 73.6\\
\rowcolor{LightCyan}
\qquad \textbf{{Ditto with Xander}}&& 9h39m & 7.3 & 68.9 & 80.5\\
\hline
\textbf{{BART}} & (139M) & 6h8m & 25.3 & 51.5 & 58.1 \\
\qquad \textbf{{Ditto with Xander}}&& 5h50m & 19.5 & 39.0 & 59.0 \\
\hline
\textbf{{CodeGen}} & (350M) & 19h22m & 54.6 & 4.4 & 5.9\\
\qquad \textbf{{Ditto with Xander}}&& 18h31m & 47.5 & 6.8 & 14.4\\
\hline
\textbf{{Microsoft Phi-1.5}} & (1.3B) & 3h49m & 34.1 & 30.5 & 42.3 \\
\qquad \textbf{{Ditto with Xander}}&& 3h41m & 18.1 & 59.8 & 72.0\\
\hline
\textbf{{ChatGPTv3.5}} & (Unknown) & N/A & 1.8 & 40.4 & 58.3 \\
\hline \\
\end{NiceTabular}

\caption{Table showcasing the results of various LMs with and without using Xander. High accuracy, low training and generation times are desirable. The best performing model is
\hl{highlighted.}
}
\label{table:generalisability}
\end{center}
\end{table*}

\subsection{Xander Ablation}
\label{sc:Xander Ablation}
For the ablation experiment we keep the experimental setup exactly the same as it is in the generalisability experiment and only consider the CodeT5 small and CodeT5 base models.

The aim of the ablation experiment is to investigate what the most important factors for determining Xander's performance are. To this end, we measure the
accuracy and runtime benefits of Normalized SQL, \QC{},
\QTR{} module, and whether examples are provided as a part of the input specification. The experimental results are reported in Table \ref{table:ablation}.
Training configurations are in bold. The word ``examples" in training configuration refers to whether or not a user-provided example was included in the task description. In normal font, we have the settings used for inference. The entry ``1 attempt" means that we stop generating queries after we find a single complete query. The entry ``multiple attempts" means we repeatedly generate queries, stopping when we find one that satisfies the user-provided example (or we reach one-minute time limit). The entry ``PQC" refers to including the \QC{} and the entry ``repair" refers to including Enumerative Repair in the \QTR{} Module.

\begin{tcolorbox}[colback=gray!7, colframe=gray!7, boxrule=0pt, left=5pt, right=5pt, top=5pt, bottom=5pt]
\textbf{The most important factor in determining Xander's accuracy is Normalized SQL which provides a 9.5\% execution accuracy increase} when a single attempt is allowed. The addition of the \QC{} gives further 4.2\% improvement to exaction accuracy. Using examples as part of the description, only improves exaction accuracy by 0.3\%. Finally, the \QTR{} module improves exaction accuracy only by 0.2\%, but improves the time to find a query by an average of 25\%.
\end{tcolorbox}

\begin{table*}
\renewcommand{\arraystretch}{1.2}
\begin{center}
\begin{NiceTabular}{l r c c c c}
Model & \shortstack{\textbf{Training Configuration} \\ Xander options}    & \shortstack{Training \\ Time} &  \shortstack{Validation Time \\ per sample (s)} & \shortstack {Validation Exact\\ Match Accuracy (\%)} & \shortstack{Validation Execution \\ Accuracy (\%)} \\
\hline
\textbf{CodeT5 small} & \textbf{SQL + No Examples + No Finetuning}& \textbf{0h0m} & - & - & -\\
& {1 attempt}  & - & 30.6 & 0 & 0 \\
\hline
\textbf{CodeT5 small*} & \textbf{SQL + No Examples}& \textbf{3h4m} & - & - & -\\
& {1 attempt}  & - & 11.6 & 1.7 & 2.4 \\
\hline
\textbf{CodeT5 small} & \textbf{SQL + No Examples} & \textbf{3h4m} & - & - & -\\
& {1 attempt}  & - & 0.5 & 46.9 & 48.7\\
\hline
\textbf{CodeT5 small } & \textbf{Normalized SQL  + No Examples} & \textbf{2h59m} & - & - & -\\
& {1 attempt}  & - & 0.9 & 56.4 & 58.4\\
\hline
\textbf{CodeT5 small } & \textbf{Normalized SQL  + Examples} & \textbf{2h59m} & - & - & -\\
& {1 attempt}  & - & 0.8 & 56.8 & 59.7\\
& {PQC  + 1 attempt}  & - & 1.1 & 59.0 & 63.9\\
& {multiple attempts}  & - & 12.8 & 66.8 & 76.0\\
& {PQC  + multiple attempts}  & - & 10.2 & 68.3 & 78.4\\
& {repair  + PQC  + multiple attempts}  & - & 9.1 & 66.0 & 78.6\\
\hline
\textbf{CodeT5 base } & \textbf{Normalized SQL  + Examples} & \textbf{9h39m} & - & - & -\\
& {PQC  + 1 attempt}  & - & 2.7 & 62.9 & 66.5\\
& {PQC  + multiple attempts}  & - & 10.3 & 71.2 & 80.4\\
\rowcolor{LightCyan}
& {repair + PQC + multiple attempts} & - & 7.3 & 68.9 & 80.5\\
\hline \\
\end{NiceTabular}

\caption{Table showcasing 
a Xander ablation study. The ``*" denotes that weights were initialised randomly as opposed to using the pre-trained weights.
The row containing the best exact match accuracy is
\hl{highlighted.}
}
\label{table:ablation}
\end{center}
\end{table*}

\subsection{\QC{} Comparison Experiment}
\label{sc:QCCompareExperiment}
The \QC{} comparison experiment aims to compare whether the \SQC{} or \NQC{} is more accurate at detecting syntax, runtime and example errors.

\begin{figure}
\begin{center}
  \includegraphics[width=0.98\linewidth]{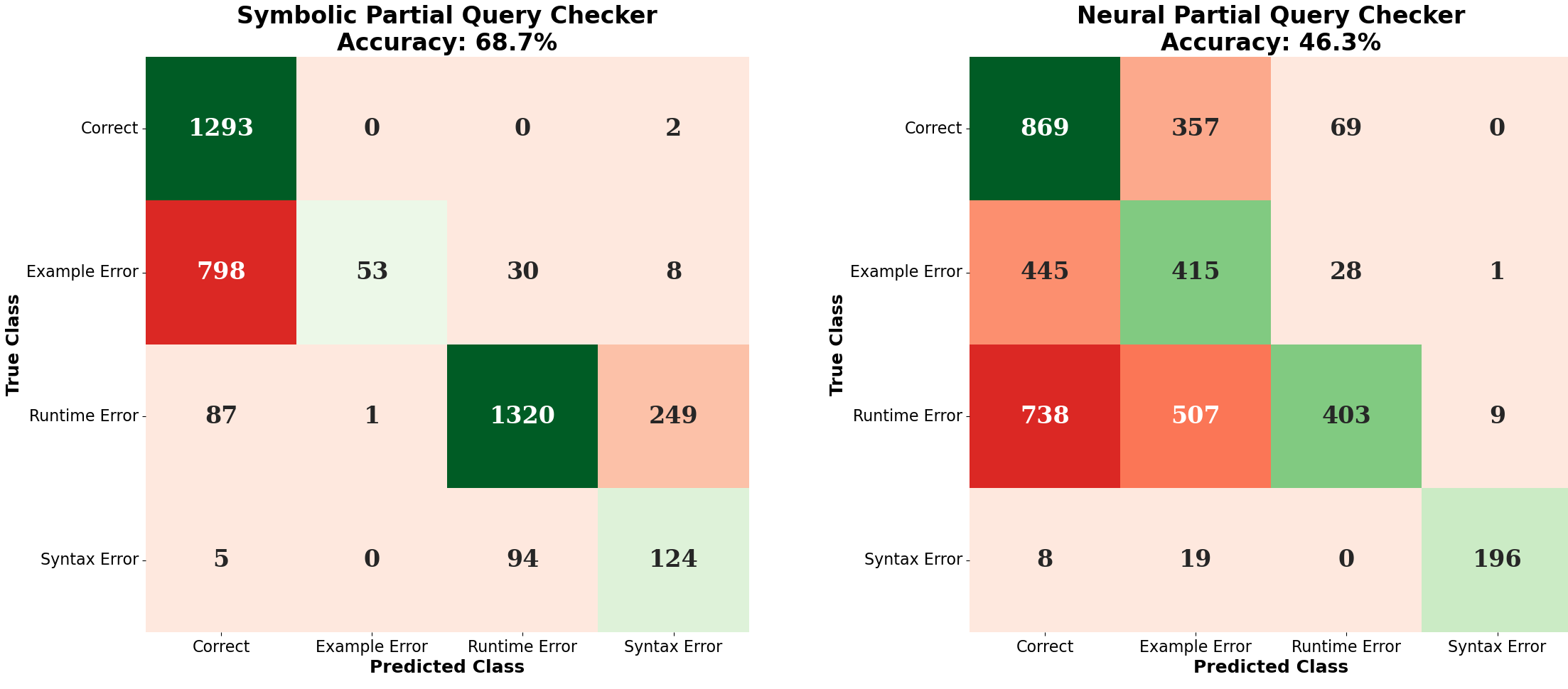}  
\end{center}
  \caption{Confusion Matrix for \SQC{} (left) and \NQC{} (right). High values in the diagonal (green) are good, high values off diagonal (red) are bad. The rows of the confusion matrix represent the true label which was obtained by executing the query. An ideal \QC{} would only have elements on the diagonal. By summing the rows, we see that most common type of mistake the Transformer Generator made was a runtime error. The columns represent the predicted class. The elements in the diagonal represent how many times predicted class matched up with the true class (higher is better). The \SQC{} makes mistakes in the error type detection due to reasons explained in Section \ref{sc:Error Type Detection}. }
\label{fig:SymbolicNeural}
\end{figure}

The \NQC{} was trained as explained in Section~\ref{sc:NQC}.
Queries from the validation set were then used to compare the efficacy of \NQC{} and \SQC{}. In particular, we generate top four most promising queries for each validation question. We then compare the two approaches of classifying them. Namely, we compare the classifications obtained by using the \SQC{} with the classification obtained by using the \NQC{}. Finally, we construct two confusion matrices to compare their accuracy (Figure~\ref{fig:SymbolicNeural}). The results allow us to draw the following conclusions:

\begin{tcolorbox}[colback=gray!7, colframe=gray!7, boxrule=0pt, left=5pt, right=5pt, top=5pt, bottom=5pt, breakable]
\begin{itemize}
\item[$\bullet$] 
Overall, \textbf{the \SQC{} is 22\% more accurate than the \NQC{} at recognising when a query contains a mistake.} This can be at least partly attributed to the \NQC{}'s inability to recognise runtime errors which occurred in 28\% of the queries. 
\item[$\bullet$] \textbf{The \SQC{} cannot differentiate between example errors and correct queries.} This is because only column types are used to detect example errors and column type mistakes are rarely ($1.3\%$ of all queries) made by the LM.

\item[$\bullet$] \textbf{The \NQC{} can detect most syntax errors.} While this is a promising finding, it is worth noting that the LM also rarely makes syntax errors. Indeed, only 5\% of the queries contain a syntax error.
\end{itemize}
\end{tcolorbox}

\subsection{Some Additional Results Explanations}
\label{sc:ResultsExplanation}
This section aims to provide an explanation for some counter-intuitive results obtained.

\paragraph*{Exact Match vs Execution Accuracy} In this work, we used execution accuracy as the main metric to measure accuracy. This is because exact-match accuracy suffers from a high false-negative rate. For example, if a database contains an integer column $\mathit{age}$, then the condition $age > 34$ would be logically the same as $age \geq 35$; however, these would be marked as different by the exact-match metric. 

\paragraph*{Error Type Detection} 
\label{sc:Error Type Detection}
The \SQC{} may misclassify queries for two main reasons. Firstly, it fails to catch example errors because only type checking is used to detect them. Secondly, sometimes the LM generates SQL instead of Normalized SQL despite being finetuned on Normalized SQL. This latter scenario was encountered twice in our experiments, leading to a correct query being tagged as having a syntax error.

\paragraph*{Runtime with One Attempt} In Xander, we found that the \QC{}
decreases runtime when multiple attempts are allowed, but increases runtime when a single attempt is allowed. This is because when a single attempt is allowed and no checking is done, we can quickly generate a query containing an error. When a \QC{} is used, we could spend up to a minute searching for a valid query.

\section{Comparison to Existing Works}
To the best of our knowledge, the only other work that accepts both NL and I/O examples is \textbf{Duoquest}~\cite{DuoQuest}. While they do not support 445 (43\%) queries of the Spider validation dataset (e.g., clauses with multiple selection predicates), for the supported queries, they report an \textbf{execution accuracy of 63.5\%}.

From text-to-SQL, 
\textbf{Picard}~\cite{Picard} is the closest to our approach, as it enhances fine-tuned LMs out-of-the-box by rejecting inadmissible tokens. However, Picard does not use I/O examples, explores the solution space differently, and doesn't attempt to repair solutions. In experiments, Picard improves \textbf{exact match accuracy by 5.15\% and execution accuracy by 6.5\%} over base LMs—less than the improvements achieved by Xander (6\% and 10.9\%, respectively). However, since Picard doesn't use I/O examples, a direct comparison is not possible.

The majority of the works on text-to-SQL require customising existing LMs. For instance, 
\textbf{RASAT}~\cite{RASAT} augments
the self-attention modules in a model's encoder and introduces new parameters to the model.
When customising T5, it achieves \textbf{exact match accuracy of 72.6\% and execution accuracy of 76.6\%} on the Spider validation set.
\textbf{REDSQL}~\cite{RESDSQL} breaks SQL generation into the generation of a skeleton of SQL keywords, which is then filled in with the missing values. 
For this purpose, it relies on a 
a ranking-enhanced encoder to alleviate the effort of the schema linking and a skeleton aware decoder to implicitly guide the SQL parsing. By customising T5-base, REDSQL achieves \textbf{exact match accuracy of 71.7\% and execution accuracy of 77.9\%} on the Spider validation set. 
While the results for RASAT and REDSQL are given as reference points, they are not directly comparable to Xander. RASAT and REDSQL require customising a base model, and do not accept I/O examples. On the other hand, Xander is designed to enhance any LM without modifications.

\paragraph{Multi-agent Architectures with LMs}
While, initially, LMs were used as black-box, monolithic entities, recently, there has been a shift towards architectures that foster some form of logical reasoning as part of the problem-solving process, sometimes by leveraging additional, possibly non-neural systems~\cite{DBLP:journals/corr/abs-2205-00445,DBLP:journals/corr/abs-2208-14271,huggingGPT,DBLP:conf/nips/Wei0SBIXCLZ22}.
Given that LLMs were shown to have difficulty with proof planning when using a linear search strategy~\cite{DBLP:conf/iclr/Saparov023},
other works are focused on decision-making and space exploration~\cite{DBLP:journals/corr/abs-2305-05364,DBLP:journals/corr/abs-2304-11477,DBLP:conf/nips/YaoYZS00N23,DBLP:journals/corr/abs-2305-08291}. 
As opposed to these works, we propose a neurosymbolic architecture for generating SQL queries that uses verification and repair modules to guide a non-linear exploration of the solution space.

\section{Threats To Validity}
While this work has presented experimental evidence that LMs augmented with symbolic reasoning techniques give a significant accuracy and runtime boost to LMs in the text-to-SQL-with examples task,
we now examine threats to the validity of these findings.

 \textbf{Symbolic reasoning may be less effective at improving huge LMs that make very few mistakes.} In particular, larger models are better at understanding the user's intent and better at generating queries that have correct syntax and satisfy the user's example. While this doesn't seem to be a problem as far as we scaled up (Xander still provides good results for Microsoft Phi-1.5), we cannot guarantee the same improvements beyond what we have tested.

\textbf{Examples are not always easy to provide.} For example, when a user's query contains an aggregation, they will not be able to specify a single tuple of the execution result. Without the example, the symbolic component's becomes less effective. This threat can be mitigated by allowing more control of the specification (e.g.\ by allowing the user to specify the output type such as number or string).

\textbf{Imperfect dataset.}
While the Spider~\cite{Spider} dataset is realistic overall and spans many domains,
we cannot guarantee the applicability of our methods beyond it.

\section{Conclusions}
This work explored whether neurosymbolic approaches could serve as a alternative to scaling model size. 
Specifically, we built a tool called Xander for the text-to-SQL-with-examples task. Xander is able to dismiss large parts of the search-space by normalizing SQL and pruning bad queries before they have finished generating. Evaluation of Xander showed that:

 \noindent\textbf{Neurosymbolic approaches led to significant improvements in runtime and accuracy.} The neurosymbolic tool Xander improves the LM's accuracy by an average of 10.9\% and generation speed by an average of 28\% on the text-to-SQL-with-examples task. 
 
 \noindent\textbf{When using Xander, 
 a smaller LM can 
    outperform its four-times-larger counterpart.}
    This means that neurosymbolic approaches provide a compelling alternative to the de facto industry standard of improving performance by scaling model size. 
 
 \noindent\textbf{Using a symbolic approach can detect errors in SQL queries where neural approaches struggle}. The \SQC{} achieves an accuracy of 69\% but the \NQC{} achieves only 46\%. This suggests that actor-critic reinforcement learning is not an appropriate choice for the text-to-SQL-with-examples task.

\bibliographystyle{ieeetr}
\bibliography{references.bib}
\end{document}